\documentclass[11pt,fleqn]{article}

\usepackage{stmaryrd,amssymb,amsmath,amsthm,url}

\newtheorem{theorem}{Theorem}
\newtheorem{lemma}{Lemma}  
\newtheorem{proposition}{Proposition}  
  
\newtheorem{corollary}{Corollary}

\newtheorem{claim}{Claim}  
\newtheorem{definition}{Definition}  
\newtheorem{example}{Example}
\theoremstyle{definition}

\newcommand{\ctpga}{\mathtt{p2pga}}
\newcommand{\length}{\ell}
\newcommand{\ANT}{\textit{TEC-AntiAUT}\,}
\newcommand{\AUT}{\textit{TEC-AUT}\,}
\newcommand{\flip}{\textit{flip}}
\newcommand{\swap}{\textit{swap}}
\newcommand{\rev}{\mathit{rev}}
\newcommand{\lef}{\unlhd}
\newcommand{\rig}{\unrhd}

\newcommand{\instr}[1]{\ensuremath{/#1}}
\newcommand{\pinstr}[1]{\ensuremath{{+}/#1}}
\newcommand{\ninstr}[1]{\ensuremath{{-}/#1}}
\newcommand{\rlinstr}[1]{\ensuremath{\backslash#1}}
\newcommand{\prlinstr}[1]{\ensuremath{{+}\backslash#1}}
\newcommand{\nrlinstr}[1]{\ensuremath{{-}\backslash#1}}
\newcommand{\jum}[1]{\ensuremath{/\##1}}
\newcommand{\bjum}[1]{\ensuremath{\backslash\##1}}
\newcommand{\ter}{\ensuremath{\:!\:}}
\newcommand{\abo}{\ensuremath{\#\:}}

\newcommand{\rlextr}[1]{\extr{#1}^{\overleftarrow{~}}}
\newcommand{\lrextr}[1]{\extr{#1}^{\overrightarrow{~}}}
\newcommand{\extr}[1]{\ensuremath{|#1|}}
\newcommand{\PI}{\ensuremath{\mathcal U}}

\newcommand{\diamarrow}[1]{\ensuremath{\mathbin{   %
                  \begin{picture}(3,2)
                  \put(.09,.1){$\langle\,#1\,\rangle$}
                  \put(.65,-.55){\vector(-1,-1){2.5}}
                  \put(2.35,-.55){\vector(1,-1){2.5}}
                  \end{picture}
                  }}}
\newcommand{\diamleftarrow}[1]{\ensuremath{\mathbin{ %
                  \begin{picture}(3,2)
                  \put(.09,.1){$\langle\,#1\,\rangle$}
                  \put(.65,-.55){\vector(-1,-1){2.5}}
                  \put(2.35,-.55){\line(1,-1){2.5}}
                  \end{picture}
                  }}}
\newcommand{\prefa}[1]{\ensuremath{\mathbin{        %
                  \begin{picture}(3,2)(-.54,.1)
                  \put(-0.3,.2){$\lbrack\,#1\,\rbrack$}
                  \put(.95,-.5){\line(0,-1){2.4}}
                  \end{picture}
                  }}}
\newcommand{\prefarrow}[1]{\ensuremath{\mathbin{    %
                  \begin{picture}(3,2)(-.54,.1)
                  \put(-0.3,.2){$\lbrack\,#1\,\rbrack$}
                  \put(.95,-.5){\vector(0,-1){2.4}}
                  \end{picture}
                  }}}

\newcommand{\di}{\mathsf{D}}
\newcommand{\st}{\mathsf{S}}

\newcommand{\NI}{{\mathcal I}}
\newcommand{\NP}{{\mathcal P}}
\newcommand{\NT}{{\mathbb T}}
\newcommand{\NZ}{{\mathbb Z}}
\newcommand{\nat}{\Nat}
\newcommand{\Nat}{{\mathbb N}}
\newcommand{\Nplus}{{\mathbb N}^+}

\newcommand{\tr}{\ensuremath{{\mathtt{true}}}}
\newcommand{\fa}{\ensuremath{{\mathtt{false}}}}

\newcommand{\BTA}{\ensuremath{\mathrm{BTA}}}
\newcommand{\BTAinf}{\ensuremath{\BTA^\infty}} 
\newcommand{\pcc}[3]{\ensuremath{#2 \unlhd #1 \unrhd #3}}
\newcommand{\less}{\ensuremath{\sqsubseteq}}
\newcommand{\Res}{\ensuremath{\mathit{Res}}}
\newcommand{\method}[1]{\ensuremath{\mathit{#1}}}
\newcommand{\fma}[2]{\ensuremath{\focus{#1}.\method{#2}}}
\newcommand{\focus}[1]{\ensuremath{\mathit{#1}}}
\newcommand{\ser}{\ensuremath{\mathcal H}} 
\newcommand{\use}[1]{\ensuremath{\mathbin{\mathalpha{/}_{\focus{#1}}}}}

\title{\titel}

\author{
	Jan A.\ Bergstra \and
	Alban Ponse \\
\\
  {\small
	  Section Software Engineering,
	  Informatics Institute,
	  University of Amsterdam}\\
	{\small URL: \url{www.science.uva.nl/~{janb/, alban/}}
	}
}
\date{}

\begin{document}
%\maketitle

\parindent=8mm

{~}
\vspace{2cm}
\begin{center}
{\Large\bf An Instruction Sequence Semigroup with Involutive Anti-Automorphisms}
\end{center}
\vspace{4mm}

\begin{center}
{\large J.A.\ BERGSTRA}\footnote{\label{fn:SSE}%
Section Theoretical Computer Science,
Informatics Institute,
University of Amsterdam.
The authors acknowledge support from
the NWO project Thread Algebra for Strategic Interleaving.
Email: {\tt \{J.A.Bergstra,A.Ponse\}@uva.nl}.}\hspace{1em}
{\large A.\ PONSE}$^{\ref{fn:SSE}}$
\end{center}
\vspace{3ex}

\date{}

\begin{abstract}
We introduce an algebra of instruction sequences by
presenting a semigroup $C$
in which programs can be represented
without directional bias:
in terms 
of the next instruction to be executed,  
$C$ has both forward and backward instructions and a
$C$-expression can be interpreted starting from any
instruction.
We provide equations for 
thread extraction, i.e., $C$'s program semantics.
Then we consider thread extraction compatible 
(anti-)homomorphisms and (anti-)automorphisms.
Finally we 
discuss some expressiveness results.
\end{abstract}

\section{Introduction}
\label{sec:intro}
In this paper three types of mathematical objects play a
basic role:
\begin{enumerate}
\item Pieces of code, i.e., finite \emph{sequences of 
instructions}, given some set $\NI$ of instructions. 
A (computer) \emph{program} is in our case 
a piece of code that
satisfies the additional property that each state of its
execution is prescribed by an instruction
(typically, there are no
jumps outside the range of instructions).
\item 
Finite and infinite sequences of \emph{primitive} 
instructions (briefly, SPIs),
the mathematical objects denoted by pieces of code (in
particular by programs). Primitive instructions are taken
from a set $\PI$ that (possibly after some renaming) is a 
strict subset of $\NI$. 
The execution of a SPI is \emph{single-pass}: 
it starts with executing the first primitive instruction, 
and each primitive instruction is dropped
after it has been executed or jumped over.
\item
\emph{Threads}, the mathematical objects representing
the execution behavior of programs and used as their
program semantics.
Threads are defined using polarized actions and a 
certain form of conditional composition.
\end{enumerate}

While each (computer) program can be considered 
as representing a sequence
of instructions, the converse is not true.
Omitting a few lines of code from a (well-formed) program 
usually results in an ill-formed program, if the remainder can
be called a program at all.
Before we discuss the instruction sequence semigroup
mentioned in the title of this paper
we briefly consider ``threads'', the mathematical objects 
representing the execution behavior of 
programs, or, more generally, of instruction sequences.
Threads as considered here resemble finite state 
schemes that represent the execution of
imperative programs in terms of their (control) actions.
We take an abstract point of view and only consider
actions and tests with symbolic names ($a,b,\ldots$):
\\{\small
\setlength{\unitlength}{1.4ex}
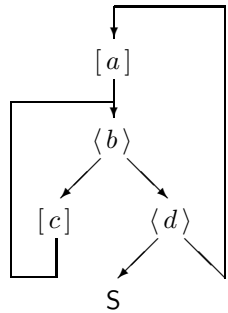
\begin{picture}(30,22)(-4,0)
\put(4,10){\diamarrow{{b}}}
\put(4,15){\prefarrow{{a}}}
\put(7.6,5){\diamleftarrow{{d}}}
\put(0.4,5){\prefa{{c}}}
\put(19,0){\begin{tabular}[b]{l}
In this picture,
\\[2mm]
$[\,a\,]$ models the execution of action $a$
and
\\ its descent
leads to the state thereafter
\\
(and likewise for $[\,c\,]$);
\\[2mm]
$\langle\,b\,\rangle$ models a the execution of test 
action
\\
$b$; its left descent
models the ``true-case''
\\
and its right one the ``false-case''
\\
(and likewise for $\langle\,d\,\rangle$);
\\[2mm]
 $\st$ is the state that models termination.
\end{tabular}}
\put(5,0){$\st$}
\put(12.45,1.95){\line(0,1){17.05}}
\put(5.5,19){\line(1,0){6.97}}
\put(5.5,19){\vector(0,-1){2}}
\put(-1,2){\line(1,0){2.9}}
\put(-1,2){\line(0,1){11}}
\put(-1,13){\line(1,0){6.5}}
\end{picture}
}
\\[4mm]
Finite state threads as the one above
can be produced in many ways,
and a primary goal of program algebra (PGA) is to study which
primitives and program notations serve that purpose well. 
The first publication on PGA is the paper~\cite{BL02}.
A basic expressiveness result states that the class of SPIs 
that can be directly represented in PGA 
(the so-called \emph{periodic} SPIs) corresponds with
these finite state threads: each PGA-program produces 
upon execution a finite state thread, 
and conversely, each finite state thread is produced by
some PGA-program.

In this paper we introduce a set of instructions that 
also suits the above-mentioned purpose well
and that at the same time has
nice mathematical properties. Together with 
concatenation|its
natural operation|it forms
a semigroup with involutions that we call $C$ (for 
``code''). A simple involutive 
anti-automorphism\footnote{We refer to \cite{Ha88} 
as a general reference for
algebraic notions.}
transforms each $C$-program into one of which the
interpretation from right to left 
produces the same thread as the original program.
Furthermore we define
some homomorphisms and automorphisms that 
preserve the threads produced by $C$-expressions,
thereby exemplifying a simple case of systematic 
program transformation.
We generalize
this approach by defining bijections on finite state
threads and describe the associated automorphisms and
anti-automorpisms on $C$, which all are
generated from simple involutions. Finally, we study
a few basic expressiveness questions about $C$.

The paper is structured as follows:
In Section~\ref{sec:threads} we review
threads in the setting of program algebra. 
Then, in Section~\ref{sec:code} we introduce the semigroup
$C$ of sequences of instructions that this paper is about.
In Section~\ref{sec:extr}
we define thread extraction on $C$, thereby giving semantics
to $C$-expressions: each $C$-expression produces a finite
state thread.
In Section~\ref{sec:exp1} we define `$C$-programs' and
show that these are sufficient to produce
finite state threads.
Furthermore, only
certain test instructions in $C$ are necessary to preserve 
$C$'s expressive power.

Section~\ref{sec:homo} is about a thread extraction
preserving homomorphism on $C$ and a related anti-homomorphism.
Then, in
Section~\ref{sec:biau} we define a natural class of 
bijections on threads and establish a relation with
a class of automorphisms on $C$, and in
Section~\ref{sec:anti} we do the same thing with respect
to a related class of anti-automorphisms on $C$.

In
Section~\ref{sec:exp2} we further consider $C$'s 
instructions in the perspective of 
expressiveness and show
that restricting to a bound on the
counters of jump instructions yields a loss in expressive power.
In Section~\ref{sec:br} we use Boolean registers to 
facilitate easy programming of finite state threads, and in
Section~\ref{sec:description} we relate the length of a
$C$-program to 
the number of states of the thread it produces.

In Section~\ref{Conc} we discuss $C$ as a 
context in which some fundamental questions about programming
can be further investigated and come up with some conclusions. 

The paper is ended with an
appendix that contains some background information
(Sections~\ref{app:A}, \ref{app:B} and \ref{app:C}).

\section{Basic Thread Algebra}
\label{sec:threads}

In this section we review threads as they
emerge from the behavioral
abstraction from programs.
Most of this text is taken from~\cite{PZ06}.

\bigskip

Basic Thread Algebra (\BTA) is a form of process
algebra which is tailored to the description of
sequential program behavior. 
Based on a set $A$ of \emph{actions},
it has the following constants and operators:
\begin{itemize}
\item the \emph{termination} constant $\st$,
\item the \emph{deadlock} or \emph{inaction} constant $\di$,
\item for each $a\in A$, a binary 
  \emph{postconditional composition}
  operator $\_  \unlhd a \unrhd \_$.
\end{itemize}
We use \emph{action prefixing} 
$a \circ P$ as an abbreviation for
$\pcc aPP$ and take $\circ$ to bind strongest.
Furthermore, for $n\geq 1$ 
we define $a^n\circ P$ by $a^1\circ P = a\circ P$ and 
$a^{n+1}\circ P = a\circ (a^n\circ P)$.

The operational intuition
is that each action represents a command which is to be
processed by the execution environment of the thread.
The processing of a command may involve a change of state of this 
environment.\footnote{%
	For the definition of threads we completely abstract
	from the environment.
	In Appendix~\ref{app:C} we define services which model
	(part of) the environment, and thread-service composition.
} 
At completion of the processing of the command, the 
environment
produces a reply value \tr\ or \fa.
The thread $\pcc aPQ$
proceeds as $P$ if the processing of $a$ yields \tr,
and it proceeds as $Q$ if the processing of
$a$ yields \fa.

Every thread in \BTA\ is finite in the sense that there
is a finite upper bound to the number of consecutive 
actions it can perform. 
The \emph{approximation operator} 
$\pi:\nat\times\BTA\rightarrow\BTA$ 
gives the behavior up to a specified depth.
It is defined by 
\begin{enumerate}
\item $\pi(0,P) = \di$,
\item $\pi(n+1,\st) = \st$, $\pi(n+1,\di) = \di$, 
\item $\pi(n+1,\pcc aPQ) = \pcc a {\pi(n,P)}{\pi(n,Q)}$,
\end{enumerate}
for $P,Q\in\BTA$ and $n\in\nat$.
We further write $\pi_n(P)$ instead of $\pi(n,P)$.
We find that for every $P\in\BTA$, there exists an $n\in\nat$ such that
\[ \pi_n(P)=\pi_{n+1}(P) = \cdots = P.
\]

Following the metric theory  of \cite{BZ82} in the form developed 
as the basis of the introduction of processes in \cite{BK84}, 
\BTA\ has a completion \BTAinf\ 
which comprises also the infinite threads.
Standard properties of the completion technique yield that we may take 
\BTAinf\ as the cpo consisting of all so-called
\emph{projective} sequences:\footnote{%
	The cpo is based on the partial ordering $\less$ 
	defined by $\di\less P$, and $P\less P'$, $Q\less Q'$ 
	implies $\pcc aPQ\less \pcc a{P'}{Q'}$.
}
\[
  \BTAinf = \{ (P_n)_{n\in \nat} \mid
  \forall n \in \nat\  (P_n\in\BTA \ \&\ \pi_n(P_{n+1})=P_n) \}.
\]
For a detailed account of this construction see~\cite{BB03}
or~\cite{V05}. On \BTAinf, equality is defined componentwise:
$(P_n)_{n\in \nat}=(Q_n)_{n\in \nat}$ 
if for all $n\in\nat$, $P_n=Q_n$.

Overloading notation, 
we now define
the constants and operators of \BTA\ 
on \BTAinf:
\begin{enumerate}
\item\label{item:A} $\di=(\di,\di,\ldots)$ and $\st=(\di,\st,\st,\ldots)$;
\item $\pcc a{(P_n)_{n\in\nat}}{(Q_n)_{n\in\nat}} = (R_n)_{n\in\nat}$
with\label{item:B} $\begin{cases}
R_0=\di,\\
R_{n+1} = \pcc a {P_n}{Q_n}.\end{cases}$
\end{enumerate}
The elements of \BTA\ are included in \BTAinf\
by a mapping following this definition. E.g., 
\[a\circ\st\mapsto (P_n)_{n\in\nat}\quad\text{with $P_0=\di$,
$P_1=a\circ\di$ and for $n\geq 2$, $P_n=a\circ \st$}.\]
It is not difficult to show that the projective sequence of 
$P\in\BTA$ thus defined equals $(\pi_n(P))_{n\in\nat}$.
We further use this inclusion of finite threads
in \BTAinf\ implicitly and write $P,Q,\ldots$ to denote
elements of \BTAinf.

We define the set $\Res(P)$ of \emph{residual threads}
of $P$ inductively as follows:
\begin{enumerate}
\item $P\in\Res(P)$, 
\item $\pcc aQR\in\Res(P)$ implies $Q\in\Res(P)$ and $R\in\Res(P)$.
\end{enumerate}
A residual thread may be reached (depending on the
execution environment) by performing zero or more actions.
A thread $P$ is \emph{regular} if $\Res(P)$ is finite.
Regular threads are also called \emph{finite state threads}.

A \emph{finite linear recursive specification} 
over \BTAinf\ is a set of equations
\[ x_i = t_i
\] 
for $i\in I$ with $I$ some finite index set, 
variables $x_i$, and all $t_i$ terms of the form
$\st$, $\di$, or $x_j\unlhd a\unrhd x_k$ with $j,k\in I$.
Finite linear recursive specifications represent
continuous operators having unique fixed points~\cite{V05}.

\begin{theorem}\label{thm:regthread}
For all $P\in\BTAinf$,
$P$ is regular iff $P$ is the solution of
a finite linear recursive specification.
\end{theorem}
\begin{proof}
Suppose $P$ is regular.
Then $\Res(P)$ is finite, so $P$ has
residual threads $P_1,\ldots, P_n$ with $P=P_1$.
We construct a finite linear recursive specification
with variables $x_1,\ldots,x_n$ as follows:
\[
  x_i = 
  \begin{cases}
  \di &\text{if } P_i=\di,\\
  \st &\text{if } P_i=\st,\\
  \pcc{a}{x_j}{x_k} &\text{if }P_i=\pcc a{P_j}{P_k}.
  \end{cases}
\] 

For the converse, assume that
$P$ is the solution of some finite linear recursive 
specification $E$ with variables $x_1,\ldots,x_n$.
Because the variables in $E$
have unique fixed points, we know that there are
threads $P_1,\ldots, P_n\in\BTAinf$ with $P=P_1$,
and for every $i\in\{1,\ldots,n\}$,
either 
$P_i=\di$, $P_i=\st$, or
$P_i = \pcc{a}{P_j}{P_k}$
for some $j,k\in\{1,\ldots,n\}$.
We find that $Q\in\Res(P)$ iff 
$Q=P_i$ for some $i\in\{1,\ldots,n\}$.
So $\Res(P)$ is finite, and $P$ is regular.
\end{proof}

\begin{example}\label{exproc}
The regular threads 
$a^2\circ \di$ and 
$a^\infty = a\circ a\circ \cdots$
are the respective fixed points for $x_1$ in the 
finite linear recursive specifications 
\begin{enumerate}
\item
	$\{x_1=a\circ x_2,~x_{2}=a\circ x_{3},~x_{3}=\di\}$, 

\item
	$\{x_1=a\circ x_{1}\}$.
\end{enumerate}
\end{example}

In reasoning with finite linear recursive specifications, we
shall often
identify variables and their fixed points. For example,
we say that $P$ \emph{is} the thread defined by $P=a\circ P$
instead of stating that $P$ equals the fixed point for $x$ 
in the finite linear recursive specification $x=a\circ x$.
In this paper we write 
\[\NT_{reg}\]
for the set of regular threads in \BTAinf.

An elegant result based on \cite{BH}
is that equality of recursively specified
regular threads can be easily decided.
Because one can always
take the disjoint union of two finite linear recursive specifications
it suffices to consider a 
single finite linear recursive specification 
$\{P_i=t_i\mid 1\leq i\leq n\}$. Then $P_i =P_j$ 
follows from
$\pi_{n-1}(P_i)=\pi_{n-1}(P_j)$.
Thus, 
it is sufficient to decide whether
two certain finite threads are equal. In Appendix~\ref{app:B}
we provide a proof sketch.

\section{\textit{C}, a Semigroup for Code}
\label{sec:code}
In this section we introduce the sequences of instructions that
form the main subject of this paper.
We call these sequences ``pieces of
code'' and use the letter $C$ to represent the resulting
semigroup. The set $A$ of actions represents a 
parameter for $C$ (as it does for $\BTA$).

\bigskip

For $a\in A$ and $k$ ranging over $\Nplus$ 
(i.e., $\Nat\setminus\{0\}$), $C$-expressions
are of the following form:
\begin{eqnarray*}
P&::=&\instr a~\Big|~\pinstr a~\Big|~ \ninstr a~\Big|~ \jum k~\Big|~
\rlinstr a~\Big|~\prlinstr a~\Big|~ \nrlinstr a~\Big|~ 
 \bjum k~\Big|~ 
\ter ~\Big|~\abo ~\Big|~ P;P
\end{eqnarray*}
In $C$ the operation ``;''
is called \emph{concatenation} and 
all other syntactical categories are called \emph{$C$-instructions}:
\begin{description}
\item[$\instr a$] is a \emph{forward basic instruction}. It 
prescribes to perform action $a$
and then (irrespective of the Boolean reply) to execute the instruction 
concatenated to its right-hand side; if there is no such instruction,
deadlock follows. 
\item[$\pinstr a$] and $\ninstr a$ are 
\emph{forward test instructions}.
The \emph{positive} forward test instruction $\pinstr a$
prescribes to perform action $a$
and upon reply $\tr$ to execute the instruction 
concatenated to its right-hand side, 
and upon reply $\fa$ to execute the second instruction 
concatenated to its righthand side; if there is no such 
instruction to be executed, deadlock follows.
For the \emph{negative} forward test instruction 
$\ninstr a$, execution 
of the next instruction is prescribed by the complementary replies.
\item[$\jum k$] is a \emph{forward jump instruction}. It 
prescribes to execute the instruction that is $k$ positions
to the right and deadlock if there is no such instruction. 
\item[$\rlinstr a,~\prlinstr a,$] $\nrlinstr a$ and $\bjum k$
are the \emph{backward} versions of the instructions mentioned
above. For these instructions, orientation is from right to left.
For example, $\rlinstr a$ prescribes to perform action $a$
and then to execute the instruction 
concatenated to its left-hand side; if there is no such instruction,
deadlock follows. 
\item[$\ter$] is the \emph{termination instruction} and
prescribes successful termination.
\item[$\abo$] is the \emph{abort instruction} and
prescribes deadlock.
\end{description}

\bigskip

\noindent
For $C$ there is one axiom:
\begin{align}
\label{A}
(X;Y);Z&=X;(Y;Z).
\end{align}
By this axiom, $C$ is a semigroup and we shall not 
use brackets in repeated concatenations.
As an example, 
\[\pinstr a;\ter;\bjum2\]
is considered an appropriate $C$-expression. 
The instructions for termination and deadlock
are the only instructions that do not specify further 
control of execution.

Perhaps the most striking aspect of $C$ is that its sequences
of instructions have no directional bias.
Although most program notations have a left to right
(and top to bottom) natural order, symmetry arguments
clarify that an orientation in the other direction
might be present as well. 

It is an empirical fact that imperative program 
notations in the vast majority of cases make use
of a default direction, inherited from the
natural language in which a program notation is naturally
embedded. This embedding is caused by the language
designers, or by the language that according to the
language designers will be the dominant mother tongue 
of envisaged programmers. None of these matters can
be considered core issues in computer science. 

The fact, however, that imperative programs invariably
show a default directional bias itself might admit 
an explanation in terms of complexity of design,
expression or execution, and $C$ provides a context in
which this advantage may be investigated.

Thus, in spite of an overwhelming evidence of the 
presence of directional bias in `practice' we
propose that the primary notation for sequences of
instructions to be used for theoretical work is
$C$ which refutes this bias.
Obviously, from $C$ one may derive a dialect $C'$ by writing
$a$ for $\instr a$, $+a$ for $\pinstr a$, $-a$ for
$\ninstr a$ and $\# k$ for $\jum k$. Now there is 
a directional bias and in terms of bytes, the
instructions are shorter.
As explained in Section~\ref{sec:exp1}, the instructions
$\rlinstr a$, $\prlinstr a$ and $\nrlinstr a$ can be eliminated,
thus obtaining a smaller instruction set which is more easily 
parsed. One may also do away with $a$ and $-a$ in favor of $+a$,
again reducing the number of instructions.
Reduction of the number of instructions leads to longer
sequences, however, and where the optimum of this
trade off is found is a matter which lies outside the
theory of instruction sequences per se.
We further discuss the nature of $C$ in Section~\ref{Conc}.

\section{Thread Extraction and \textit{C}-Expressions}
\label{sec:extr}
In this section we
define thread extraction on $C$. For
a $C$-expression $X$,
$\lrextr  X$
denotes the thread produced by $X$ when execution
started at the leftmost or ``first'' instruction,
thus $\lrextr{..}$ is an operator that assigns a thread
to a $C$-expression. We prove that this is always
a regular thread.
We also consider right-to-left thread extraction
where thread extraction starts at the righmost
of a $C$-expression.

\bigskip

We will use auxiliary functions
$\extr X_j$ with $j$ ranging over the integers $\NZ$
and we define
\begin{align*}
\lrextr  X&=\extr  X_1,
\end{align*}
meaning that thread extraction starts at 
the first (or leftmost) instruction of $X$. 
For $j\in\NZ$, $\extr X_j$ is defined in 
Table~\ref{tab:extr}.

\begin{table}[htb]
\hrule
\begin{align}
\nonumber
&\text{Let  
$X=i_1;\ldots;i_n$ and $j\in\NZ$.}\\[2mm]
\nonumber
&\text{For $j\in\{1,\ldots, n\}$, }\\
\nonumber
&\quad\extr{X}_j=
\begin{cases}
a\circ\extr{X}_{j+1}&\text{if $i_j=\instr a$},\\
 \extr{X}_{j+1} \unlhd
a\unrhd\extr{X}_{j+2}&\text{if $i_j= \pinstr a$},\\
 \extr{X}_{j+2} \unlhd
a\unrhd\extr{X}_{j+1}&\text{if $i_j= \ninstr a$},\\
\extr{X}_{j+k}&\text{if $i_{j}=\jum k$},
\\[4mm]
a\circ\extr{X}_{j-1}&\text{if $i_{j}=\rlinstr a$},\\
 \extr{X}_{j-1} \unlhd
a\unrhd\extr{X}_{j-2}&\text{if $i_j= \prlinstr a$},\\
 \extr{X}_{j-2} \unlhd
a\unrhd\extr{X}_{j-1}&\text{if $i_j= \nrlinstr a$},\\
\extr{X}_{j-k}&\text{if $i_{j}=\bjum k$},
\\[4mm]
\st&\text{if $i_{j}=\ter$},\\
\di&\text{if $i_{j}=\abo$},
\end{cases}
\\[4mm]
\label{1}
&\text{for $j\not\in\{1,\ldots, n\}$, }\quad\extr{X}_{j}=\di.
\end{align}
\hrule
\caption{Equations for thread extraction}
\label{tab:extr}
\end{table}

A special case arises if these equations applied from left to right
define a loop without any actions, as in
\begin{align*}
\extr{\jum2;\instr a;\bjum 2}_1
&=\extr{\jum2;\instr a;\bjum 2}_3\\
&=\extr{\jum2;\instr a;\bjum 2}_1.
\end{align*}
For this case we have the following rule:
\begin{align}
\label{d-rule}
&\text{If the equations in Table~\ref{tab:extr} applied from left to right
yield}\\
\nonumber
&\text{a loop without any actions the extracted thread is $\di$. }
\end{align}
Rule~\eqref{d-rule} applies if and only if a 
loop in a thread extraction is
the result of consecutive jumps to jump
instructions. 

In the following we show that 
thread extraction on $C$-expressions
produces regular threads.
For a $C$-expression $X$ we define
$\length(X)\in\Nplus$ to be the length of $X$, i.e., 
its number of instructions. 

\begin{theorem}
\label{la:exp}
If $X$ is a $C$-expression 
and $i\in\NZ$,
then $\extr X_i$ 
defines a regular thread.
\end{theorem}

\begin{proof}
Assume $X$ is a $C$-expression with $\length(X)=n$.
If $i\not\in\{1,\ldots,n\}$, then $\extr X_i=\di$
by rule~\eqref{1}. In the other case, a
single application of the matching equation
in Table~\ref{tab:extr} determines for each 
$i\in\{1,\ldots,n\}$ an equation of the form
\begin{equation}
\label{aha}
\extr X_i=\extr X_j\lef a\rig\extr X_k, 
\quad\text{or }
\extr X_i=\extr X_j,
\quad\text{or }
\extr X_i=\di, 
\quad\text{or }
\extr X_i=\st
\end{equation}
where by rule~\eqref{1} we may assume that 
all expressions $\extr X_j$ and $\extr X_k$ 
occurring in the
right-hand sides satisfy $j,k\in\{1,\ldots,n\}$
(otherwise they are replaced by $\di$). 
We construct $n$ linear equations $x_i=t_i$
with the property 
that $\extr X_i$ as given by the rules for 
thread extraction is a fixed point for $x_i$:
\begin{enumerate}
\item
Define $x_i=t_i$ from
\eqref{aha} by replacing each $\extr X_j$
by $x_j$.
\item
Determine with Rule~\eqref{d-rule} all equations 
$\extr X_i=\extr X_j$ that define a loop
without actions, and replace all associated
equations $x_i=x_j$ by 
\[x_i=\di.\]
\item
Replace any remaining equation of the form 
$x_i=x_j$ by 
\[x_i=t_j\]
where 
$t_j$ is the right-hand side of the equation for
$x_j$. Repeating this procedure exhaustively yields
a finite linear specification with variables
 $x_1,\ldots,x_n$.
\end{enumerate} 
For each $i\in\{1,\ldots,n\}$
the thread defined by thread extraction on
$\extr X_i$ is a fixed point
for $x_i$.
Hence $\lrextr X$
is a regular thread, and so is $\rlextr X$.
\end{proof}

Given some $C$-expression $X$, we shall often use
$\extr X_i$  as the identifier of 
the thread defined by $\extr X_i$  
as meant in Theorem~\ref{la:exp}, 
and similar for $\lrextr X$.
As an example of thread extraction, 
consider the $C$-expression
\begin{equation}
\label{oho}
X=\instr a;\pinstr b;\rlinstr c;\pinstr d;\ter;\bjum5
\end{equation}
It is not hard to check that
$X$ produces the regular thread 
$P_1$ (i.e., $\lrextr X=P_1$)
defined by\footnote{This regular thread
$P_1$ can be visualized as was done in 
Section~\ref{sec:intro}.}
\begin{align*}
P_1&= a\circ P_2\\
P_2&= P_3\unlhd  b\unrhd P_4\\
P_3&=c\circ P_2\\
P_4&=P_5\unlhd  d\unrhd P_1\\
P_5&=\st
\end{align*}
Thread extraction defines
an equivalence on $C$-expressions,
say $X\equiv_{\rightarrow}Y$ if $\lrextr X=\lrextr Y$, 
that is not a congruence, 
e.g.,
\[{\abo}\equiv_{\rightarrow}{\jum1}\quad\text{but}\quad 
{\abo;\instr a}\not\equiv_{\rightarrow}{\jum1;\instr a}.\]

\bigskip

We define right-to-left thread extraction, notation
\[\rlextr X,\]
as the thread extraction that starts from 
the rightmost position of a piece of code:
\begin{align*}
\rlextr  X&=\extr  X_{\length(X)}
\end{align*}
where $\length(X)\in\Nplus$ is the length of $X$, i.e., 
its number of instructions. Taking $X$ as defined in 
Example~\eqref{oho}, we find 
$\rlextr X=\lrextr X$ because for that particular $X$,
$\extr X_6=\extr X_1$.
Right-to-left thread extraction also defines
an equivalence on $C$-expressions,
say $X\equiv_{\leftarrow}Y$ if $\rlextr X=\rlextr Y$, 
that is not a congruence, 
e.g.,
\[{\abo}\equiv_{\leftarrow}{\bjum1}\quad\text{but}\quad
{\instr a;\abo}\not\equiv_{\leftarrow}{\instr a;\bjum1}.\]

\section{Expressiveness of \textit{C}-Programs}
\label{sec:exp1}
In this section we introduce
the notion of a `$C$-program'.
Furthermore we
discuss a basic expressiveness result: we show that
each regular thread is the
thread extraction of some $C$-program.
Finally
we establish that we do not need all of $C$'s instructions
to preserve expressiveness.

\bigskip

\begin{definition}
\label{paria}
A \textbf{C-program}
is a piece of code $X=i_1;\ldots;i_n$ with $n~>~0$ such that 
the computation of $\extr X_j$ for each $j=1,\dots,n$
does not use equation~\eqref{1}.
In other words, there are no
jumps outside the range of $X$ and execution 
can only end by executing
either the termination instruction $\ter$
or the abort instruction $\abo$. 
\end{definition}

In the setting of program algebra we explicitly
distinguished in \cite{BP08} a ``program'' from an
instruction sequence (or a piece 
of code) in the sense that a program has
a natural and preferred semantics, while this is not the case
for the latter one. Observe that if $X$ and $Y$ are
$C$-programs, then so is $X;Y$.
A piece of code that is not a program
can be called a \emph{program fragment} because it can 
be extended to a program that yields the same thread extraction.
This follows from the next proposition, which states that
position numbers can be relativized.

\begin{proposition}
\label{prop1}
For $k\in\Nat$ and $X$ a $C$-expression,
\begin{enumerate}
\item
$\extr  X_k=\extr{\abo;X}_{k+1}$,
\item
$\extr  X_k=\extr{X;\abo}_{k}$.
\end{enumerate}
Moreover, in the case that
$X$ is a $C$-program and $1\leq k\leq\length(X)$,
\begin{itemize}
\item[3.]
$\extr X_k=\lrextr{\jum k;X}$,
\item[4.]
$\extr X_k=\rlextr{X;\bjum\length(X)+1-k}$.\end{itemize}
\end{proposition}
With properties 1 and 2 we find for example
\begin{align*}
\lrextr{\pinstr a;\bjum2}&=\extr{\pinstr a;\bjum2}_1\\
&=\extr{\abo;\pinstr a;\bjum2;\abo}_2,
\end{align*}
and since the latter piece of code is a $C$-program,
we find with property 3 another one that produces
the same thread with left-to-right thread extraction:
\[\extr{\abo;\pinstr a;\bjum2;\abo}_2=
\lrextr{\jum2;\abo;\pinstr a;\bjum2;\abo}.\]
Of course, for property 3 to be valid
it is crucial that $X$ is a $C$-program:
for example
\begin{align*}
\lrextr{\pinstr a;\bjum2}&=\extr{\pinstr a;\bjum2}_1\\
&\neq\lrextr{\jum1;\pinstr a;\bjum2}. 
\end{align*}
A similar example contradicting property~4 for 
$X$ not a $C$-program is easily found.

\begin{theorem}
\label{thm:exp}
Each regular thread in $\NT_\text{reg}$
is produced by a $C$-program.
\end{theorem}

\begin{proof}
Assume that a regular thread $P_1$ is specified
by linear equations $P_1=t_1,\ldots, P_n=t_n$.
We transform each equation into a piece of $C$-code:
\begin{align*}
P_i=\st&\mapsto \ter;\abo;\abo,\\
P_i=\di&\mapsto \abo;\abo;\abo,\\
P_i=P_j\lef a\rig P_k&\mapsto 
\begin{cases}
\pinstr a;\jum p;\jum q&\text{if }p,q>0,\\
\pinstr a;\jum p;\bjum (-q)&\text{if }p>0,~q<0,\\
\pinstr a;\bjum(-p);\jum q&\text{if }p< 0,~q>0,\\
\pinstr a;\bjum (-p);\bjum (-q)&\text{if }p,q< 0,\\
\end{cases}
\end{align*}
where $p=3(j-i)-1$ and $q=3(k-i)-2$ 
(so $p,q\in\NZ\setminus\{0\}$). 
Concatenating these pieces of code in the order
given by $P_1,\ldots,P_n$
yields a $C$-expression $X$ with $\lrextr X=P_1$.
% (and $P_i=\extr X_3(i-1)+1$). 
By construction $X$ contains no
jumps outside the range of instructions and 
therefore $X$ is a $C$-program. 
Finally, note that the instructions of $X$ are in the set
%the instruction set
$\{\pinstr a,\jum k,\bjum k,\ter,\abo\mid a\in A,~k\in\Nplus\}$.
\end{proof}

From the proof of Theorem~\ref{thm:exp} we infer
that only positive forward test instructions, jumps and 
termination are needed to
preserve $C$'s expressiveness:

\begin{corollary}
\label{cor:A}
Let $C^-$ be defined by allowing
only instructions from the set 
\[\{\pinstr a,\jum k,\bjum k,\ter\mid a\in A,~k\in\Nplus\}.
\]
Then each regular thread in $\NT_\text{reg}$
can be produced by a program in $C^-$.
\end{corollary}
\begin{proof} 
With $\abo$ added to the instruction set mentioned,
the result follows immediately from the proof of 
Theorem~\ref{thm:exp}. The use of $\abo$ in that proof
can easily be avoided, for example by setting
\begin{align*}
P_i=\st&\mapsto \ter;\jum1;\bjum1
&\quad\hfill\text{(instead of $\ter;\abo;\abo$)},\\
P_i=\di&\mapsto \jum1;\jum1;\bjum1
&\quad\hfill\text{(instead of $\abo;\abo;\abo$)}.
\end{align*}
The resulting expression clearly contains 
no jumps outside its range and is hence
a $C$-program.
\end{proof}

\section{Thread Extraction Preserving Homomorphisms}
\label{sec:homo}
In this section we consider functions on $C$  
that preserve thread extraction. We start
with a homomorphism 
that turns all basic and test instructions into their
forward counterparts, and another one that only yields
positive forward test instructions. Then we consider
an anti-homomorphism
that relates extraction with
right-to-left thread extraction.
So, these functions are very basic examples of 
program transformation.

\bigskip

Let the function $h:C\rightarrow C$ be defined on 
$C$-instructions as follows:
\begin{align*}
\instr a&\mapsto \instr a;\jum2;\abo,\\
\pinstr a&\mapsto \pinstr a;\jum2;\jum4,\\
\ninstr a&\mapsto \ninstr a;\jum2;\jum4,\\
\jum k&\mapsto \jum 3k;\abo;\abo,\\[2mm]
\rlinstr a&\mapsto \instr a;\bjum4;\abo,\\
\prlinstr a&\mapsto \pinstr a;\bjum4;\bjum8,\\
\nrlinstr a&\mapsto \ninstr a;\bjum4;\bjum8,\\
\bjum k&\mapsto \bjum 3k;\abo;\abo,\\[2mm]
\ter&\mapsto \ter;\abo;\abo,\\
\abo&\mapsto \abo;\abo;\abo.
\end{align*}
So, $h$ replaces all basic and test instructions
by fragments containing only their forward counterparts.
Defining 
\[h(X;Y)=h(X);h(Y)\]
makes
$h$ an injective homomorphism (a `monomorphism')
that preserves the equivalence obtained
by (left-to-right) thread extraction, i.e.,
\[\lrextr X=\lrextr{h(X)}.\]
This follows from the more general property
\[\extr X_{j+1}=\extr{h(X)}_{3j+1}\]
for all $j<\length(X)$, which is easy to prove by
case distinction. So,
$\lrextr X=\lrextr{h^k(X)}$, and, moreover, if $X$
is a $C$-program, then so is $h^k(X)$.

Of course many variants of the homomorphism $h$ satisfy 
the latter two properties. A particular one is the
homomorphism obtained 
from $h$
by replacement with the
following defining clauses: 
\begin{align*}
\instr a&\mapsto \pinstr a;\jum2;\jum1,\\
\ninstr a&\mapsto \pinstr a;\jum5;\jum1,\\
\rlinstr a&\mapsto \pinstr a;\bjum4;\bjum5,\\
\nrlinstr a&\mapsto \pinstr a;\bjum4;\bjum8,
\end{align*}
because now only forward positive test instructions
occur in the homomorphic image. 
In other words: with respect to thread extraction,
$C$'s expressive power is preserved if its set of instructions
is reduced to 
\[
\{\pinstr a,\jum k,\bjum k,\ter,\abo\mid a\in A,~k\in\Nplus\}.
\]
This is the syntactic counterpart of 
Corollary~\ref{cor:A}
in Section~\ref{sec:exp1}.

Let $g: C\rightarrow C$ be defined on 
$C$-instructions as follows:
\begin{align*}
\instr a&\mapsto \abo;\bjum2;\rlinstr a,\\
\pinstr a&\mapsto \bjum4;\bjum2;\prlinstr a,\\
\ninstr a&\mapsto \bjum4;\bjum2;\nrlinstr a,\\
\jum k&\mapsto \abo;\abo;\bjum 3k,\\[2mm]
\rlinstr a&\mapsto \abo;\jum4;\rlinstr a,\\
\prlinstr a&\mapsto \jum8;\jum4;\prlinstr a,\\
\nrlinstr a&\mapsto \jum8;\jum4;\nrlinstr a,\\
\bjum k&\mapsto \abo;\abo;\jum 3k,\\[2mm]
\ter&\mapsto \abo;\abo;\ter,\\
\abo&\mapsto \abo;\abo;\abo.
\end{align*}
So, $g$ replaces all basic and test instructions
by $C$-fragments containing only their backward counterparts.
Defining $g(X;Y)=g(Y);g(X)$ makes
$g$ an \emph{anti-homomorphism} that satisfies
\[\lrextr X=\rlextr{g(X)}.\]
This follows from a more general property
discussed in Section~\ref{sec:anti}.

\section{Structural Bijections and TEC-Automorphisms}
\label{sec:biau}
In this section we define \emph{structural bijections} on 
the finite state threads over $A$ as a natural type 
of (bijective) thread transformations. 
We then describe and analyze
the associated class of automorphisms on $C$, which 
appear to be generated from simple involutions. 

\bigskip

Given a bijection $\phi$ on $A$ (thus a permutation of $A$)
and a partitioning
of $A$ in $A_\tr$ and $A_\fa$, we
extend $\phi$ to a \emph{structural bijection}
on $\BTA$ by 
defining for all $a\in A$ and $P,Q\in\BTA$,
\begin{align*}
\phi(\di)&=\di,\\
\phi(\st)&=\st,\\
\phi(P\lef a\rig Q)&=\begin{cases}
\phi(P)\lef \phi(a)\rig \phi(Q)&\text{if $\phi(a)\in A_\tr$},\\
\phi(Q)\lef \phi(a)\rig \phi(P)&\text{if $\phi(a)\in A_\fa$}.
\end{cases}
\end{align*}

Structural bijections naturally extend to $\NT_\text{reg}$:
if $P_i$ is a fixed point for $x_i$ in the finite linear 
specification
$\{x_i=t_i(\overline x)\mid i=1,\ldots,n\}$, then 
$\phi(P_i)$ is a fixed point for $y_i$ in 
\begin{equation}
\label{trans}
\{y_i=\phi(t_i(\overline x))\mid 
i=1,\ldots,n,~\phi(x_i)=y_i\}.
\end{equation}
As an example, assume that $\phi(a)=b\in A_\fa$
and thread $P$ is given by
\[
P=P\lef a\rig Q,~Q=\di\]
then $P'=\phi(P)$ is defined by
\[
P'=Q'\lef b\rig P',~Q'=\di.\]

\begin{theorem}
There are $2^{|A|}\cdot |A|!$ structural bijections on 
$\BTA$, and 
thus on $\NT_\text{reg}$.
\end{theorem}

\begin{proof}
Trivial: if $|A|=n$, there are $2^n$ different
partitionings in $A_\tr$ and $A_\fa$,
and $n!$ different bijections on $A$.
\end{proof}

Each structural bijection 
can be written as the composition of
a (possibly empty) series of 
transpositions or `swaps' (its permutation part) and a 
(possibly empty) series of postconditional `flips' that model
the $\fa$-part of its partitioning. So, for a fixed $\phi$
there exist $k$ and $m$ such that
\[\phi=\overline\flip_{c_1}\circ\ldots\circ\overline\flip_{c_m}\circ
\overline\swap_{a_1,b_1}\circ\ldots\circ\overline\swap_{a_k,b_k}\]
where $\overline\swap_{a,b}$ models the 
exchange of actions $a$ and $b$, 
and $\overline\flip_c$ the postconditional flips
for $A_\fa=\{c_1,\ldots,c_m\}$,
and $\phi$ is the identity if $k=m=0$. More precisely,
\begin{align*}
&\overline\swap_{a,b}(P\lef c\rig Q)=\\
&\overline\swap_{a,b}(P)\lef \overline c\rig\overline\swap_{a,b}(Q)
\quad\text{with}\quad
\begin{cases}
\overline c=b&\text{if }c=a,\\
\overline c=a&\text{if }c=b,\\
\overline c=c&\text{otherwise},
\end{cases}
\end{align*}
and
\[
\overline\flip_c(P\lef a\rig Q)=
\begin{cases}
\overline\flip_c(Q)\lef a\rig\overline\flip_c(P)&\text{if }a=c,\\
\overline\flip_c(P)\lef a\rig\overline\flip_c(Q)&\text{otherwise}.
\end{cases}
\]
For $A=\{a_1,\ldots, a_n\}$ we can do 
with $n-1$ swaps $\overline\swap_{a_1,a_j}$ ($1<j\leq n$)
as these define any other swap by $\overline\swap_{a_i,a_j}
=\overline\swap_{a_1,a_j}\circ\overline\swap_{a_1,a_i}
\circ\overline\swap_{a_1,a_j}$, and $n$ flips 
$\overline\flip_{a_i}$ ($1\leq i\leq n$). 

We show that structural bijections naturally
correspond with a certain class of
automorphisms on $C$.

\begin{definition}
\label{TEC}
An automorphism $\alpha$ on $C$ is 
\textbf{thread extraction compatible} (TEC)
if there exists a structural bijection $\beta$ such
that the following diagram commutes:
\[\begin{array}{ccc}
C&\stackrel{\lrextr{-}}\longrightarrow&\NT_\text{reg}\\[2mm]
\phantom{\alpha}\downarrow\alpha&&\phantom{\beta}
\downarrow\beta\\[2mm]
C&\stackrel{\lrextr{-}}\longrightarrow&\NT_\text{reg}\\
\end{array}
\]
\end{definition}

\begin{theorem}
\label{thm:1}
The TEC-automorphisms on $C$ are generated by
\begin{align*}
\swap_{a,b}:&\quad\text{exchanges $a$ and $b$ in all 
instructions containing $a$ or $b$},
\\
\flip_a:&\quad\text{exchanges $+$ and $-$ in all 
test instructions containing $a$},
\end{align*}
where $a$ and $b$ range over $A$.
\end{theorem}

\begin{proof}
First we have to show that if $\alpha$ is generated
from $\swap_{a,b}$ and
$\flip_a$ ($a,b\in A$), then $\alpha$ is a TEC-automorphism.
This follows from the fact that the diagram in 
Definition~\ref{TEC} commutes for $\swap_{a,b}$ if we take 
$\beta=\overline\swap_{a,b}$ and for $\flip_a$ 
if we take $\beta=\overline\flip_a$. We show this below.

Then we have to show that if $\alpha$ is a TEC-automorphism,
then $\alpha$ is generated from swaps and flips.
Above we argued that each structural bijection
can be characterized by zero or more $\overline\swap_{a,b}$
and $\overline\flip_a$
applications. So, again it suffices to argue that for 
$\beta=\overline\swap_{a,b}$,
the diagram 
commutes if $\alpha=\swap_{a,b}$ and 
for $\beta=\overline\flip_c$ %the diagram 
if $\alpha=\flip_c$.
The general case
follows from repeated applications. 

Let $X\in C$. First
assume $\beta=\overline\flip_c$.
Following the construction in the proof of
Theorem~\ref{la:exp} we find a  
finite linear specification $\{x_i=t_i\mid i=1,\ldots,n\}$
with $n=\length(X)$ such 
that $\extr X_i$ is a fixed point for $x_i$. 
Transforming
this specification according to~\eqref{trans} 
with $\phi=\overline\flip_c$ yields 
$\{y_i=\overline\flip_c(t_i(\overline x))\mid 
i=1,\ldots,n,~\overline\flip_c(x_i)=y_i\}$. Now  
$\extr{\flip_c(X)}_i$ 
is a fixed point for $y_i$:
this also follows from the construction in the proof of 
Theorem~\ref{la:exp} and the fact that $\flip_c$ only
changes the sign of $\pm\instr c$ and $\pm\rlinstr c$
in $X$.

We now show that
$\overline\flip_c(\extr X_i)$ is a fixed
point for $y_i$ by a case distinction
on the form of $t_i$ in the equations 
$x_i=t_i$ ($i=1,\ldots,n$):
\begin{itemize}
\item
If $x_i= x_j\lef c\rig x_k$ then 
$\extr X_i=\extr X_j\lef c\rig \extr X_k$, so
\begin{align*}
\overline\flip_c(\extr X_i)&=
\overline\flip_c(\extr X_{j}\lef c\rig \extr X_{k})
\\&= \overline
\flip_c(\extr X_{k})\lef c\rig \overline\flip_c(\extr X_{j}).
\end{align*}
Note that in this case $y_i=y_k\lef c\rig y_j$.
\item 
If $x_i= x_j\lef a\rig x_k$ with $a\neq c$, then
$\extr X_i=\extr X_j\lef a\rig \extr X_k$, so
\begin{align*}
\overline\flip_c(\extr X_i)&=
\overline\flip_c(\extr X_{j}\lef a\rig \extr X_{k})
\\&= \overline
\flip_c(\extr X_{j})\lef a\rig \overline\flip_c(\extr X_{k}).
\end{align*}
Note that in this case $y_i=y_j\lef a\rig y_k$.
\item If $x_i= \st$, then $\extr X_i=\st$
and $y_i=\st$. Also
$\overline\flip_c(\extr X_i)=\st$.
\item If $t_i= \di$, then $\extr X_i=\di$ and
$y_i=\di$. Also
$\overline\flip_c(\extr X_i)=\di$.
\end{itemize}
So in all cases $\overline\flip_c(\extr X_i)$ 
is a fixed point for $y_i$. Hence, 
$\extr{\flip_c(X)}_i=\overline\flip_c(\extr X_i)$
and thus $\lrextr{\flip_c(X)}=\overline\flip_c(\lrextr X)$.

In a similar way it follows that 
$\extr{\swap_{a,b}(X)}_i=\overline\swap_{a,b}(\extr X_i)$.
\end{proof}

Note that $\swap_{a,a}$ is the identity and so is
$\flip_a\circ\flip_a$. 
Furthermore, for $a\neq b$ we have $\swap_{a,b}=\swap_{b,a}$ and 
\[\swap_{a,b}\circ\flip_{c}=
\begin{cases}
\flip_{c}\circ\swap_{a,b}&\text{if $c\not\in\{a,b\}$},\\
\flip_{d}\circ\swap_{a,b}&\text{if $\{a,b\}=\{c,d\}$}.
\end{cases}\]
This implies that each TEC-automorphism can be 
represented as
\[\flip_{c_1}\circ\ldots\circ\flip_{c_m}\circ
\swap_{a_1,b_1}\circ\ldots\circ\swap_{a_k,b_k}.\]
Similarly as remarked above,
for $A=\{a_1,\ldots, a_n\}$ we can do 
with $n-1$ swaps $\swap_{a_1,a_j}$ ($1<j\leq n$)
as these define any other swap.

We further write $\AUT$ for the set of TEC-automorphisms, and
we say that $\swap_{a,b}$ and the structural
bijection $\overline\swap_{a,b}$ are \emph{associated}, and similar
for $\flip_a$ and $\overline\flip_a$. So, the above
result states that for the associated pair
$\alpha\in\AUT$ and structural bijection
$\overline\alpha$ the following
diagram commutes:
\[\begin{array}{ccc}
C&\stackrel{\lrextr{-}}\longrightarrow&\NT_\text{reg}\\[2mm]
\phantom{\alpha}\downarrow\alpha&&\phantom{\alpha}
\downarrow\overline\alpha\\[2mm]
C&\stackrel{\lrextr{-}}\longrightarrow&\NT_\text{reg}\\
\end{array}
\]
The following corollary of Theorem~\ref{thm:1} follows immediately.
\begin{corollary}
\label{COR1}
If $\alpha\in\AUT$, then 
$\alpha$ preserves the orientation of all
instructions and
$\alpha(i)=i\quad 
\text{for}\quad
i\in \{\jum k,\bjum k,\ter,\abo\mid k\in\Nplus\}$.
Furthermore, for each $a\in A$, $\alpha$ is determined by its
value on one of the possible four test instructions. If for
example $\alpha(\pinstr a)=\ninstr b$, then
$\alpha(\instr a)=\instr b$,
$\alpha(\ninstr a)=\pinstr b$, and the remaining
identities are given by replacing all forward
slashes by backward slashes. 
\end{corollary}

Each element $\alpha\in\AUT$ that satisfies
$\alpha^2(u)=u$ for all $C$-instructions $u$ is an 
\emph{involution}, i.e. 
\[\alpha^2(X)=X.\]
Obvious examples of involutions are $\swap_{a,b}$
and $\flip_c$, and a counter-example is
\[\alpha=\flip_b\circ\swap_{a,b}\]
because
\[\alpha^2=
\flip_b\circ\swap_{a,b}\circ\flip_b\circ\swap_{a,b}=
\flip_b\circ\flip_a\circ\swap_{a,b}\circ\swap_{a,b}=
\flip_b\circ\flip_a.\]
However, $\alpha^2$ is an involution (because compositions 
of flip commute).

\section{TEC-Anti-Automorphisms}
\label{sec:anti}
In this section we consider the relation between 
structural bijections on threads and an associated
class of
anti-automorphisms on $C$. Recall that a function 
$\phi$ is an anti-homomorphism
if it satisfies $\phi(X;Y)=\phi(Y);\phi(X)$. 
Furthermore, we show how the monomorphism $h$ defined in 
Section~\ref{sec:homo} is systematically related to the 
anti-homomorphism $g$ defined in that section.

\bigskip

Define the {anti-automorphism} $\rev: C\rightarrow C$ 
(reverse) on $C$-instructions by the exchange of all forward 
and backward orientations:
\begin{align*}
\instr a&\mapsto \rlinstr a,\\
\pinstr a&\mapsto \prlinstr a,\\
\ninstr a&\mapsto \nrlinstr a,\\
\jum k&\mapsto \bjum k,\\[2mm]
\rlinstr a&\mapsto \instr a,\\
\prlinstr a&\mapsto \pinstr a,\\
\nrlinstr a&\mapsto \ninstr a,\\
\bjum k&\mapsto \jum k,\\[2mm]
\ter&\mapsto\ter,\\
\abo&\mapsto\abo.
\end{align*} 
Then $\rev^2(X)=X$, so $\rev$ is an {involution}. 
Furthermore, it is immediately clear
that for all $X\in C$,
\[\lrextr{X}=\rlextr{\rev(X)}.\]

\begin{definition}
\label{anti}
An anti-automorphism $\alpha$ on $C$ is 
\textbf{thread extraction compatible} (TEC)
if there exists a structural bijection $\beta$ such
that the following diagram commutes:
\[\begin{array}{ccc}
C&\stackrel{\lrextr{-}}\longrightarrow&\NT_\text{reg}\\[2mm]
\phantom{\alpha}\downarrow\alpha&&\phantom{\beta}
\downarrow\beta\\[2mm]
C&\stackrel{\rlextr{-}}\longrightarrow&\NT_\text{reg}\\
\end{array}
\]
\end{definition}

We write $\ANT$ for the set of thread extraction
compatible anti-automorphisms on $C$. The following result 
establishes a strong connection between $\AUT$ and $\ANT$.

\begin{theorem} 
$\ANT=\{\rev\circ\alpha\mid\alpha\in\AUT\}$.
\end{theorem}

\begin{proof}

Let $\gamma\in\ANT$, so $\gamma$ is an anti-automorphism and 
there is a structural bijection $\beta$ such that
$\rlextr{\gamma(X)}=\beta(\lrextr X)$ for all $X$.
By Theorem~\ref{thm:1}, $\beta=\overline\alpha$
for some $\alpha\in\AUT$ and 
$\beta(\lrextr X)=\lrextr{\alpha(X)}$ for all $X$,
and thus 
\begin{equation}
\label{refute}
\rlextr{\gamma(X)}=\rlextr{\rev\circ\alpha(X)}
\quad\text{for all $X$}.
\end{equation}
This defines $\gamma$ on $\{\jum k,\bjum k,\ter,\abo\mid
k\in\Nplus\}$.
By Corollary~\ref{COR1}, $\alpha$ is
determined by its definition on all positive
forward test instructions. So, if
for $a,b\in A$,
$\alpha(\pinstr a)=\pm\instr b$ then we
find by~\eqref{refute} with
$X=\pinstr a;\ter$ that
$\gamma(\nrlinstr a)=\mp\rlinstr b$. Since $\alpha$
is determined for all other instructions
containing $a$, also $\gamma$ is fully determined 
for all instructions containing $a$.
It follows that 
$\gamma=\rev\circ\alpha$, thus 
$\gamma\in \{\rev\circ\alpha\mid\alpha\in\AUT\}$. 

Conversely, if 
$\gamma\in \{\rev\circ\alpha\mid\alpha\in\AUT\}$, 
say $\gamma=\rev\circ\alpha$ 
with $\alpha\in\AUT$, 
then $\rlextr{\gamma(X)}=\lrextr{\alpha(X)}=
\beta(\lrextr X)$ for some structural bijection $\beta$ and
all $X$.  Furthermore, $\gamma$ is an anti-automorphism, so
$\gamma\in\ANT$.
\end{proof}

Observe that for all $\alpha\in\AUT$,
\(\alpha\circ\rev=\rev\circ\alpha\)
and for all  $\alpha,\beta\in\ANT$,
\(\alpha\circ\beta\in\AUT.\)
Using the notation for associated pairs we find for
$\beta=\rev\circ\alpha\in\ANT$ that the following diagram commutes:
\[\begin{array}{ccc}
C&\stackrel{\lrextr{-}}\longrightarrow&\NT_\text{reg}\\[2mm]
\phantom{\beta}\downarrow\beta&&\phantom{\alpha}
\downarrow\overline\alpha\\[2mm]
C&\stackrel{\rlextr{-}}\longrightarrow&\NT_\text{reg}\\
\end{array}
\]
Note that we use $\overline\alpha$, i.e.,
the associated structural 
bijection of $\alpha$, in this diagram. 

\bigskip

Another application with $\rev$ is the following:
for ${h}: C\rightarrow C$ a monomorphism,
the following diagram commutes:
\[\begin{array}{ccc}
C&\stackrel{\rev\circ h}\longrightarrow&C\\[2mm]
\phantom{h}\downarrow h&&\phantom{\stackrel{\rlextr{-}}{}}
\downarrow\stackrel{\rlextr{-}}{}\\[2mm]
C&\stackrel{\lrextr{-}}\longrightarrow&\NT_\text{reg}\\
\end{array}
\]
As an example, consider the anti-homomorphism $g$ defined 
in Section~\ref{sec:homo}: indeed $g=\rev\circ h$ for 
the homomorphism $h$ defined in that section.

\section{Expressiveness and reduced instruction sets}
\label{sec:exp2}
In this section we further consider $C$'s
instructions in the perspective of expressiveness.
We show that
setting a bound on the size of 
jump counters in $C$ does have consequences
with respect to expressiveness:
let 
\[C_k\]
be defined by allowing only jump instructions 
with counter value $k$ or less.

\bigskip

We first introduce some auxiliary notions:
following the definition of residual threads
in Section~\ref{sec:threads}, we say that
thread $Q$ is a \emph{$0$-residual}
of thread $P$ if $P=Q$, and an
\emph{$n+1$-residual} of $P$ if for some $a\in A$,
$P=\pcc a{P_1}{P_2}$ and $Q$ is an $n$-residual of $P_1$
or of $P_2$. 
Note that a finite thread (in \BTA) only
has $n$-residuals for finitely many $n$, 
while for the thread
$P$ defined by $P=a\circ P$ it holds that $P$ is an 
$n$-residual of
itself for each $n\in\Nat$.

Let $a\in A$ be fixed and $n\in\Nplus$. 
Thread $P$ has the \emph{$a$-$n$-property} if
$\pi_n(P)=a^n\circ \di$ and $P$ has $2^n-1$ (different)
$n$-residuals which all have a first approximation not equal 
to $a\circ\di$. So, if a thread $P$ has the 
{$a$-$n$-property}, then
$n$ consecutive $a$-actions can be executed and each 
sequence of $n$ replies leads
to a unique $n$-residual. Moreover, none of these
residual threads starts with an $a$-action (by the
requirement on their first approximation).
We note that for each $n\in\Nplus$ we can find a finite
thread with the $a$-$n$-property. In the next section we
return to this point.

A piece of code $X$ has the {$a$-$n$-property}
if for some $i$, 
$\extr X_i$ has this property.
It is not hard to see that in this case
$X$ contains at least $2^n-1$ different $a$-tests.
As an example, consider 
\[X=\ter;\rlinstr b;\prlinstr a;\pinstr a;\bjum 2;
\pinstr a;\jum2; \instr c;\abo\]
Clearly, $X$ has the $a$-2-property because
$\extr X_4$ has this property: its 2-residuals are
$b\circ\st$, $\st$, $\di$ and $c\circ\di$, so each
thread is not equal to one of the others and does
not start with an $a$-action.

\bigskip

Note that if a piece of code $X$ has the 
{$a$-$(n+k)$-property}, then it also has the $a$-$n$-property.
In the example above, $X$ has the 
$a$-1-property because $\extr X_3$ has this property
(and $\extr X_6$ too).

\begin{lemma}
\label{a-n-prop}
For each $k\in\Nat$ there exists $n\in\Nplus$ 
such that no $X\in C_k$
has the $a$-$n$-property. 
\end{lemma}

\begin{proof}
Suppose the contrary and let $k$ be minimal in this respect.
Assume for each $n\in\Nplus$,
$Y_n\in C_k$ has the $a$-$n$-property.

Let $B=\{\tr,\fa\}$. For $\alpha,\beta\in B^*$
we write
\[\alpha\preceq\beta\]
if $\alpha$ is a prefix of $\beta$, and we write
$\alpha\prec\beta$ or $\beta\succ\alpha$
if $\alpha\preceq\beta$ and $\alpha\neq\beta$.
Furthermore,
let 
\[B^{\leq n}=\bigcup_{i=0}^n B^i,\]
thus $B^{\leq n}$ contains all $B^*$-sequences 
$\alpha$ with 
$\length(\alpha)\leq n$ (there are 
$2^{n+1}-1$ such sequences).

Let $g:\Nat\rightarrow \Nat$ be such that 
$\extr{Y_n}_{g(n)}$ has the $a$-$n$-property.
Define
\[f_n: B^{\leq n}\rightarrow\Nplus\]
by $f_n(\alpha)=m$ if the instruction
reached in $Y_n$ when execution started at position
$g(n)$ after the replies to $a$ according to
$\alpha$ 
has position $m$.
Clearly, $f_n$ is an injective function.

In the following claim we show that under the supposition
made in this proof a certain form of squeezing holds:
if $k'$ is sufficiently large, then for all $n>0$
there exist 
$\alpha,\beta,\gamma\in B^{k'}$ with 
$f_{k'+n}(\alpha)<f_{k'+n}(\beta)<f_{k'+n}(\gamma)$
with the property that 
$f_{k'+n}(\alpha)<f_{k'+n}(\beta')<f_{k'+n}(\gamma)$ 
for each extension $\beta'$ of $\beta$ within 
$B^{\leq k'+n}$.
This claim is proved by showing that not having this
property implies that ``too many'' such extensions 
$\beta'$ exist. 
Using this claim 
it is not hard to contradict the minimality of $k$.

\begin{claim}
\label{claimB}
Let $k'$ satisfy $2^{k'}\geq2k+3$. Then for all $n>0$
there exist $\alpha,\beta,\gamma\in B^{k'}$ 
with \[f_{k'+n}(\alpha)<f_{k'+n}(\beta)<f_{k'+n}(\gamma)\]
such that for 
each extension $\beta'\succeq\beta$ in 
$B^{\leq k'+n}$, 
\[f_{k'+n}(\alpha)<f_{k'+n}(\beta')<f_{k'+n}(\gamma).\]
\end{claim}

\begin{proof}[Proof of Claim~\ref{claimB}.]
Let $k'$ satisfy $2^{k'}\geq2k+3$. 
Towards a contradiction, suppose the stated
claim is not true for some $n>0$.
The sequences in $B^{k'}$ 
are totally ordered by $f_{k'+n}$, say 
\[f_{k'+n}(\alpha_1)<f_{k'+n}(\alpha_2)< \ldots
<f_{k'+n}(\alpha_{2^{k'}}).\]
Consider the following list of sequences:
\begin{align*}
\alpha_1,\underbrace{\alpha_2,\dots,\alpha_{2k+2}},
\alpha_{2k+3}\\
\text{choices for }\beta\hspace{10mm}
\end{align*}
By supposition there is 
for each choice $\beta
\in\{\alpha_2,\ldots,\alpha_{2k+2}\}$ 
an 
extension $\beta'\succ\beta$ in $B^{\leq {k'+n}}$
with
\[\text{either}\quad
f_{k'+n}(\beta')<f_{k'+n}(\alpha_1),\quad\text{or}
\quad
f_{k'+n}(\beta')>f_{k'+n}(\alpha_{2k+3}).
 \]
Because there are
$2k+1$ choices for $\beta$, assume
that at least $k+1$
elements $\beta\in\{\alpha_2,
\ldots,\alpha_{2k+2}\}$ 
have an extension $\beta'$ with 
\[f_{k'+n}(\beta')<f_{k'+n}(\alpha_1)\]
(the assumption $f_{k'+n}(\beta')>
f_{k'+n}(\alpha_{2k+3})$ for at
least $k+1$ elements $\beta$ with extension $\beta'$
leads to a similar argument). 
Then we obtain a contradiction 
with respect to $f_{k'+n}$:
for each of the sequences $\beta$ in the subset just
selected
and its extension $\beta'$,
\[f_{k'+n}(\beta')<f_{k'+n}(\alpha_1)<f_{k'+n}(\beta),\] 
and there are at least
$k+1$ different such  pairs $\beta,\beta'$ 
(recall $f_{k'+n}$ is injective). But
this is not possible with jumps of at
most $k$ because the $f_{k'+n}$ values of
each of these pairs 
define a path in $Y_{k'+n}$
that never has a gap
that exceeds $k$ and
that passes position $f_{k'+n}(\alpha_1)$, while 
different paths never share a position.
This finishes the proof of Claim~\ref{claimB}.
\end{proof}

Take according to 
Claim~\ref{claimB} an appropriate 
value $k'$, some value $n>0$ and 
$\alpha,\beta,\gamma\in B^{k'}$.
Consider ${Y_{k'+n}}$
and mark the positions that are used for
the computations according to $\alpha$ and $\gamma$:
these computations both start in position $g({k'+n})$ 
and end in
$f_{k'+n}(\alpha)$ and $f_{k'+n}(\gamma)$, respectively.
Note that the set of marked positions never has a gap
that exceeds $k$.

Now consider a computation that starts from instruction
$f_{k'+n}(\beta)$ in $Y_{k'+n}$, a position in 
between $f_{k'+n}(\alpha)$
and $f_{k'+n}(\gamma)$. By Claim~\ref{claimB}, the first
$n$ $a$-instructions have positions in between 
$f_{k'+n}(\alpha)$
and $f_{k'+n}(\gamma)$ and none of these are marked.
Leaving out all marked positions and adjusting the
associated jumps yields a piece of code, say $Y$, with 
smaller jumps, thus in $C_{k-1}$, that has the 
$a$-$n$-property.
Because $n$ was chosen arbitrarily, this contradicts 
the initial supposition that $k$ was minimal.
\end{proof}

\begin{theorem}
\label{thm:2}
For any $k\in\Nplus$, not all threads 
in $\BTA$ can be expressed in $C_k$. This 
is also the case if 
thread extraction may start at arbitrary
positions.
\end{theorem}

\begin{proof}
Fix some value $k$. Then, by
Lemma~\ref{a-n-prop} we can find a value $n$
such that no $X\in C_k$ has the $a$-$n$-property.
But we can define a finite thread that has this property.
\end{proof}

In the next section we discuss a systematic 
approach to define finite threads that have the 
$a$-$n$-property.

\section{Boolean Registers for Producing Threads}
\label{sec:br}
In this section we briefly discuss the use of Boolean
registers to ease programming in $C$. 
This is an example of so-called \emph{thread-service}
composition. In appendix~\ref{app:C} we provide
a brief but general introduction to {thread-service}
composition.

\bigskip

Consider \emph{Boolean registers} named 
$b1,b2,\ldots,bn$ which all are initially set to $F$ (false)
and can be set to $T$ (true). 
We write $bi(b)$ with $b\in\{T,F\}$ to indicate that $bi$'s
value is $b$.
The action $bi.set{:}b$ sets register $bi$ to $b$ and 
yields \tr\ as its reply. The action $bi.get$ 
reads the value 
from register $bi$ and provides this value as its reply. 
The defining rules for threads in \BTA\ that use 
one of these
registers are for $b,b'\in\{T,F\}$,
$i\in\{1,\ldots,n\}$:
\begin{align*}
  \st\use{bi}bi(b) &= \st,\\
  \di\use{bi}bi(b) &= \di, \\ 
  (\pcc{bi.set{:}b'}PQ)\use{bi}bi(b) &=
 P\use{bi}bi(b'),\\
  (\pcc{bi.get}PQ)\use{bi}bi(b) &=
  \begin{cases}
 P\use{bi}bi(b)&
 \quad\text{if }b=T,\\
 Q\use{bi}bi(b)&
 \quad\text{if }b=F,
 \end{cases}
 \end{align*}
and, if none of these rules apply, 
 \[
 (\pcc{a}PQ)\use{bi}bi(b) =
 \pcc{a}{(P\use{bi}bi(b))}{(Q\use{bi}bi(b))}.\]
The operator $\use {bi}$ is called
the \emph{use operator} and stems from~\cite{BP02}.
Observe that the requests to the service $bi$
do not occur as actions in
the behavior of a thread-service composition. 
So the composition hides the associated actions.

\bigskip

As a simple example consider the $C$-program $X$
that has extra instructions based on the set
$\{bi.set{:}b, bi.get\mid b\in\{T,F\},~i\in\{1,2\}\}$:
\begin{align*}
X=~
&\pinstr a;\instr b1.set{:}T;\\
&\pinstr a;\instr b2.set{:}T;\\
&\pinstr b1.get;c;d;\\
&\pinstr b2.get;c;d;\ter
\end{align*}
Then one can derive (recall the initial value of
$b1$ and $b2$ is $F$):
\begin{align*}
(\lrextr X\use{b1}b1)\use{b2}b2
&=(\pcc a{\extr X_3\use{b1}b1(T)}{\extr X_3\use{b1}b1(F)})\use{b2}b2\\
&=\pcc a{(\pcc a{R_1}{R_2})}{(\pcc a{R_3}{R_4})}
\end{align*}
where $R_1=c\circ d\circ c\circ d\circ\st$
(case $T,T$), 
$R_2=c\circ d\circ d\circ\st$
(case $T,F$), 
$R_3=d\circ c\circ d\circ\st$
(case $F,T$), and
$R_4=d\circ  d\circ\st$ (case $F,F$). So, the
four possible
combinations of the values of $b1$ and $b2$ yield the
different 2-residuals $R_1, \ldots,R_4$.
Clearly, $X$ has the $a$-2-property.
The particular form of the
$C$-program $X$ already suggests
how to generalize $X$
to a family of $C$-programs $Z_n$ ($n\in\Nplus$)
such that 
\[((\lrextr{Z_n}\use{b1}b1)...)\use{bn} bn\]
has the $a$-$n$-property:
\begin{align*}
Z_n=
&\pinstr a;~\instr b1.set{:}T;\\
&\pinstr a;~\instr b2.set{:}T;\\
&\ldots\\
&\pinstr a;~\instr bn.set{:}T;\\
&\pinstr b1.get;~c;d;\\
&\pinstr b2.get;~c;d;\\
&\ldots\\
&\pinstr bn.get;~c;d;\ter
\end{align*}
Each series of $n$ replies to the positive testinstructions
$\pinstr a$ has a unique continuation after which $Z_n$ terminates 
successfully: the number of $\tr$-replies matches the number 
of $c$-actions, and their ordering that of the occurring 
$d$-actions. 
Obviously, each thread 
$((\lrextr{Z_n}\use{b1}b1)...)\use{bn} bn$ 
is a finite thread in \BTA\ and can thus be produced 
by a $C$-program not using Boolean registers
(cf.\ Theorem~\ref{thm:exp}). 

More information about thread-service composition is
given in Appendix~\ref{app:C}.

\section{On the Length of \textit{C}-Programs for Producing  Threads}
\label{sec:description}
$C$-programs can be viewed as descriptions of 
finite state threads. 
In this section we
consider the question which program length is needed
to produce a finite state thread.
We also consider the case that auxiliary Boolean registers 
are used for producing threads, which can
be a very convenient feature as was shown in the previous
section.
We find upper and lower bounds for the lengths of
$C$-programs.

\bigskip

For $k,n\in\Nplus$ let
\[\psi(k,n)\in\Nplus\]
be the minimal value such that each thread over alphabet 
$a_1,\ldots,a_k$ with at most $n$ states can be expressed
as a $C$-program with at most $\psi(k,n)$ 
instructions. Furthermore, let
\[\psi_{br}(k,n)\in\Nplus\]
be the minimal value such that each thread over alphabet 
$a_1,\ldots,a_k$ with at most $n$ states can be expressed
as a $C$-program with at most $\psi_{br}(k,n)$ 
instructions including those to use Boolean registers.

It is not hard to see that
\[\psi(k,n)\leq 3n\quad\text{and}\quad
\psi_{br}(k,n)\leq 3n\]
because each state can be described by either the
piece of code
\[\pinstr a_i;u;v\]
with $u$ and $v$ jumps to the pieces of code that model the
two successor states, or by $\ter$ or $\abo$.
Presumably, a sharper {upper bound} for both $\psi(k,n)$ and
$\psi_{br}(k,n)$ can be found.

As for a {lower bound} for $\psi_{br}(k,n)$, 
we can use auxiliary Boolean registers by forward basic 
instructions
\begin{eqnarray*} 
\instr bi.set{:}T\\
\instr bi.set{:}F\\
\instr bi.get
\end{eqnarray*} 
and their backward and test counterparts. So,
each Boolean register $bi$ comes with
18 different instructions, and of course at most 
$\psi_{br}(k,n)$ of these can be used.

Programs containing at most $l=\psi_{br}(k,n)$ instructions,
contain per position $i$
at most $l-1$ jump instructions, namely 
jumps to all other (at most $l-1$) positions
in the program. 

So, if we restrict to $k=1$, say $\instr a$ is the 
only forward basic instruction involved 
(with backward and test variants yielding 5
more instructions) and include the termination
instruction $\ter$ and the abort instruction $\abo$, the 
admissible instruction alphabet counts
\[
2+6+(l-1)+ 18l
\]
instructions. Because $l\geq 1$, this is bounded by
$26l$ instructions, and therefore we count
\[(26l)^l\]
syntactically different programs.

A lower bound on the number of threads with $n$ states 
over one action $a$
can be estimated as follows:
let $F$ range over all functions
\[
\{1,\ldots,n-1\}\mapsto \{0,1,\ldots,n-1\},\]
thus there are $n^{n-1}$ different $F$.
Define threads $P^F_k$ for $k=0,\ldots,n-1$ by
\begin{align*} 
P^F_0&=\st\\
P^F_{i+1}&=P^F_{F(i+1)}\unlhd a \unrhd P^F_i
\end{align*} 
We claim that for a fixed $n$ the threads $P^F_{n-1}$ 
(each one containing $n$ states $P^F_0,\ldots,P^F_{n-1}$),
are for each $F$ different,
thus yielding $n^{n-1}$ different threads, so we find
\begin{equation}
\label{AA}
(26l)^l\geq n^{n-1}.
\end{equation}
Assume $n\geq 2$, thus $26\leq 25n$, thus 
$n\leq 26n -26$, thus $\displaystyle\frac n{26}\leq n-1$. 
Suppose $l<\displaystyle \frac n{26}$, then
$26l< n$ and $l< n-1$, which contradicts \eqref{AA}.
Thus 
\[l\geq \frac n{26}.\]
So, for $k=1$ and in fact for arbitrary $k\geq 1$ we find 
\[\frac n {26}\leq \psi_{br}(k,n)\leq 3n.\]

In the case that we do not allow the use of
auxiliary Boolean registers,
it follows in a same manner as above
that for arbitrary $k\geq 1$, 
\[\frac n {8}\leq \psi(k,n)\leq 3n.\]

We see it as a challenging problem to improve 
the bounds of $\psi_{br}(k,n)$ and $\psi(k,n)$.

\section{Discussion}
\label{Conc}
In this paper we proposed an algebra of instruction
sequences based on a set of instructions without
directional bias.
The use of the phrase ``instruction sequence'' asks for
some rigorous motivation. 
This is a subtle matter which defeats many common sense
intuitions regarding the science of computer programming.

The Latin source of the word `instruction' tells us no more
than that the instruction is part of a listing. On that 
basis, instruction sequence is a pleonasm and justification
is problematic.\footnote{\cite{C18}:
INSTRUCTION, in Latin 
\emph{instructio},
comes from \emph{in} and \emph{struo} to dispose or regulate,
signifying the thing laid down.

The following is taken from 
\url{http://www.etymonline.com/}. INSTRUCTION:
from O.Fr. instruction, from L. instructionem (nom. instructio) 
``building, arrangement, teaching," from instructus, pp. of 
instruere ``arrange, inform, teach," from in- ``on" + 
struere ``to pile, build" (see structure).}
We need to add the additional connotation of instruction
as a ``unit of command''. This puts instructions at a 
core position. 
Maurer's paper
\emph{A theory of computer instructions}~\cite{M06} 
provides a theory of instructions
which can be taken on board in an attempt to define what
is an instruction in this more narrow sense. Now Maurer's 
instructions
certainly qualify as such but his survey is not exhaustive.
His theory has an intentional focus on transformation of
data while leaving change of control unexplained. We hold 
that Maurer's theory, including his ongoing work on this 
theme in~\cite{M08}, provides a candidate definition for
so-called basic instructions.

At this stage different arguments can be used to make progress.
Suppose a collection $\NI$ is claimed to constitute a set
of instructions:
\begin{enumerate}
\item 
\label{one}
If the mnemonics of elements of $\NI$ are reminding
of known instructions of some low level program notations,
and if the semantics provided complies with that view,
the use of these terms may be considered justified.
\item 
If, however, unknown, uncommon or even novel instructions
are included in $\NI$, the argument of \ref{one} can not be used.
Of course some similarity of explanation can be used to carry
the jargon beyond conventional use. At some stage, however, a 
more intrinsic justification may be needed.
\item
A different perspective emerges if one asserts that 
certain instruction
sequences constitute programs, thus considering $\NI^+$ 
(i.e., finite, non-empty sequences of instructions 
from $\NI$) one
may determine a subset $\NP\subseteq\NI^+$ of programs.
Now a sequence in $\NI^+$ qualifies as a program
if and only if it is 
in $\NP$. In the context of $C$-expressions we say that
\[\pinstr a;\bjum 10;\instr b;\pinstr c;\jum8;\ter;\ter\]
is not in $\NP$ because the jumps outside the range of 
instructions cannot be given a natural and preferred
semantics, as
opposed to $\pinstr a;\bjum 1;\ter$
and $\pinstr a;\instr b;\pinstr c;\ter;\ter$. We here state
once more that we do \emph{not} consider the empty sequence
of instructions as a program, or even as an instruction 
sequence because we have no canonical meaning or even intuition 
about such an empty sequence in this context. 

\item
\label{C1}
The next question is how to determine $\NP$. At this point 
we make use of the framework of PGA~\cite{BL02,PZ06} (for a 
brief explanation of PGA see Appendix~\ref{app:A}).
A program is a piece of data for which the preferred and
natural meaning is a ``sequence of primitive instructions'', 
abbreviated to a SPI. 
Primitive instructions are defined over some collection $A$
of basic instructions. The meaning of a program $X$ is by
definition provided by means of a projection function which
produces a SPI for $X$. Using PGA as a notation for SPIs,
the projection function can be written $\ctpga$
(``$\NP$ to PGA"). The behavior
$\extr X_\NP$ for $X\in\NP$ is given by 
\[\extr X_\NP=\extr{\ctpga(X)}\]
where thread extraction in PGA, i.e., $\extr{\ldots}$,
is supposed to be known.
\item
\label{C2}
In the particular case of $\NI$ consisting of $C$'s 
instructions,
we take for $\NP$ those instruction sequences for which
control never reaches outside the sequence. These are the 
sequences that we called $C$-programs. First we restrict
to $C$-programs composed from instructions in 
$\{\instr a,\pinstr a,\ninstr a,\jum k,\bjum k, \ter,\abo
\mid a\in A,k\in\Nplus\}$
and we define 
\[F(i_1;\ldots;i_n)=(\psi(i_1);\ldots;\psi(i_n))^\omega\]
as a ``pre-projection function'' that uses 
an auxiliary function $\psi$ on these instructions:
\begin{align*}
\psi(\instr a)&=a,\\
\psi(\pinstr a)&=+a,\\
\psi(\ninstr a)&=-a,\\
\psi(\jum k)&=\# k, \\
\psi(\bjum k)&=\# n-k, \\
\psi(\ter)&=\ter,\\
\psi(\abo)&=\#0.
\end{align*}
We can rewrite each $C$-program into this restricted form
by applying the behavior preserving
homomorphism $h$ defined in Section~\ref{sec:homo}. 
Thus our final definition of a projection can
be $\ctpga=F\circ h$. Note that many alternatives
for $h$ could have been used as well (as was already noted
in Section~\ref{sec:homo}).
\item
\label{C3}
Conversely, each PGA-program can be embedded into $C$ while 
its behavior is preserved.
For repetition free programs this embedding is defined by
the addition of forward slashes and replacing $\#0$ by 
\abo.\footnote{The instruction \abo\ already 
occured in \cite{BL00}, 
but was in \cite{BL02} replaced by $\#0$, thus admitting
a more systematic treatment of ``jumps''.}
In the other case, a 
PGA-program can be embedded into PGLB, a variant of PGA
with backward jumps and no repetition operator~\cite{BL02},
and transformation from PGLB to $C$ is trivial.
\end{enumerate} 

\noindent
In the case of $C$, 
items~\ref{C1} and \ref{C2} above
should of course be \emph{proved}, i.e., for a $C$-program $X$,
\[\lrextr X=\extr X_C\quad(=\extr{\ctpga(X)}),\]
and for item~\ref{C3} a similar requirement about the
definition of $\lrextr\ldots$ should be substantiated. 
We omit these proofs as they seem rather clear.

\subsection*{Acknowledgements}
We thank Stephan Schroevers and an anonymous
referee for their useful comments and for pointing
out some errors.

\bibliographystyle{plain}

\begin{thebibliography}{58}

\bibitem{BZ82}
J.W. de Bakker and J.I. Zucker.
\newblock Processes and the denotational semantics of concurrency.
\newblock \emph{Information and Control}, 54(1-2):70-120,1982.

\bibitem{BH}
A. Barros and T. Hou.
A constructive version of AIP revisited.
\newblock Technical report PRG0802, University of Amsterdam,
January 2008.\\
Available via {www.science.uva.nl/research/prog/publications.html}.


\bibitem{BB03}
J.A. Bergstra and I. Bethke.
\newblock Polarized process algebra and program equivalence.
\newblock In J.C.M. Baeten, J.K. Lenstra, J. Parrow, G.J. Woeginger, eds.,
{\em Proceedings of ICALP 2003}, 
LNCS 2719, pages 1-21, Springer-Verlag, 2003.

\bibitem{BB05}
J.A. Bergstra and I. Bethke. 
\newblock Polarized process algebra with reactive composition. 
\newblock \emph{Theoretical Computer Science}, 343(3):285-304, 2005.

\bibitem{BK84}
J.A. Bergstra and J.W. Klop.
\newblock Process algebra for synchronous communication.
\newblock {\em Information and Control}, 60(1-3):109-137, 1984.

\bibitem{BL00}
J.A. Bergstra and M.E. Loots.
Program algebra for component code.
\newblock \emph{Formal Aspects of Computing},
 12(1):1-17, 2000.

\bibitem{BL02}
J.A. Bergstra and M.E. Loots.
\newblock Program algebra for sequential code.
\newblock {\em Journal of Logic and Algebraic Programming},
51(2):125-156, 2002.

\bibitem{BP02}
J.A. Bergstra and A. Ponse. Combining programs and state
machines. \emph{Journal of Logic and Algebraic Programming}
51(2):175-192, 2002.

\bibitem{BP08}
J.A. Bergstra and A. Ponse.
An instruction sequence semigroup with repeaters.
\newblock arXiv:0810.1151v1 [cs.PL] at http://arxiv.org/, 2008.

\bibitem{C18}
George Crabb.
\emph{English Synonyms Explained, in Alphabetical Order: 
With Copious Illustrations and Examples Drawn from the Best 
Writers}.
\newblock 
Published by Baldwin, Cradock, 1818.
Original from the New York Public Library,
Digitized Sep 25, 2006,
904 pages.

\bibitem{Ha88}
M. Hazewinkel.
Encyclopaedia of Mathematics: an updated and annotated translation of the 
Soviet ``Mathematical Encyclopaedia''.
Springer-Verlag, 2002.
  
\bibitem{M06}
W.D. Maurer. 
A theory of computer instructions. 
\newblock \emph{Science of Computer Programming},
60:244-273, 2006.
(A shorter version of this paper was published in the
\emph{Journal of the ACM}, 13(2): 226--235, 1966.)

\bibitem{M08}
W.D. Maurer.
Partially defined computer instructions and guards. 
\newblock \emph{Science of Computer Programming},
72(3):220-239, 2008.
 
\bibitem{PZ06}
A. Ponse and M.B. van der Zwaag. 
An introduction to program and thread algebra. 
\newblock In A. Beckmann et al. (editors), 
\emph{Logical Approaches to Computational Barriers: 
Proceedings CiE 2006}, LNCS 3988, pages 445-458, Springer-Verlag, 2006.

\bibitem{V05}
T.D. Vu.
\newblock Denotational semantics for thread algebra.
\newblock \emph{Journal of Logic and Algebraic Programming},
74(2):94-111, 2008.

\end{thebibliography}

\small
\appendix
\section{PGA, a summary}
\label{app:A}
Let a set $A$ of constants with typical elements 
$a,b,c,\ldots$ be given.
PGA-programs are of the following form ($a\in A,~k\in\Nat$):
\[P ::= a\mid +a\mid -a
\mid \#k\mid\;! \mid P;P\mid P^\omega.\]
Each of the first five forms above is called a  
\emph{primitive instruction}. 
We write \PI\ for the set of primitive instructions and we 
define each element of \PI\ to be a SPI (Sequence of 
Primitive Instructions).

Finite SPIs are defined using \emph{concatenation}: if
$P$ and $Q$ are SPIs, then so is 
\[P;Q\] 
which is the SPI that lists 
$Q$'s primitive instructions right after those of $P$, and we 
take concatenation to be an \emph{associative} operator.

Periodic SPIs are defined using the repetition operator:
if $P$ is a SPI, then
\[P^\omega\]
is the SPI that repeats $P$ forever, thus $P;P;P;\ldots$.
Typical identities that relate repetition and concatenation 
of SPIs are
\[(P;P)^\omega=P^\omega 
\quad\text{and}\quad 
(P;Q)^\omega=P;(Q;P)^\omega.\]
Another typical identity is 
\[P^\omega;Q=P^\omega,\]
expressing that nothing ``can follow'' an infinite repetition.

The execution
of a SPI is \emph{single-pass}: it starts with the 
first (left-most)
instruction, and each instruction is dropped after it has been
executed or jumped over.

Equations for thread extraction on SPIs, notation
\(\extr{X},\) are the following,
where $a$ ranges over $A$, $u$ over the
primitive instructions \PI, and $k\in\Nat$:
\begin{align*}
|!| &= \st&
|!;X| &= \st\\[1mm]
|\mathtt a|&=\mathtt a\circ\di&
|\mathtt a;X| &=\mathtt a \circ |X|\\[1mm]
|\mathtt{{+}a}|&=\mathtt a\circ\di
\hspace{.5cm}&
|{+}\mathtt{a};X| &= |X|\unlhd \mathtt a \unrhd |\#2;X|\\[1mm]
|{-}\mathtt a|&=\mathtt a\circ \di&
|{-}\mathtt a;X| &= |\#2;X|\unlhd \mathtt a \unrhd |X|\\[4mm]
|\#k|&=\di&
|\#0;X| &= \di\\[1mm]
&&
|\#1;X| &= |X|\\[1mm]
&&
|\#k{+}2;u| &= \di\\[1mm]
&&
|\#k{+}2;u;X| &= |\#k{+}1;X|\\[-3mm]
\end{align*}
For more information on PGA we refer to~\cite{BL02,PZ06}.

\section{Basic Thread Algebra and Finite Approximations}
\label{app:B}
An elegant result based on \cite{BH}
is that equality of recursively specified
regular threads can be easily decided.
Because one can always
take the disjoint union of two finite linear recursive
specifications,
it suffices to consider a 
single specification $\{P_i=t_i\mid
1\leq i\leq n\}$. Then 
$P_i =P_j$
follows from
\begin{equation*}\label{lin}
\pi_{n-1}(P_i)=\pi_{n-1}(P_j).
\end{equation*} 
Thus, 
it is sufficient to decide whether
two certain finite threads are equal. We provide a 
proof sketch:

For $k\geq 0$ consider the equivalence relation
$\cong_k$ on $\{P_1,\ldots,P_n\}$ defined by $P_i\cong_k P_j$ if
$\pi_k(P_i)=\pi_k(P_j)$. Then 
\begin{equation}
\label{lon}
\cong_0\;\supseteq\;\cong_1
\;\supseteq\;\cong_2\;\supseteq\ldots
\end{equation}
If
$\cong_k\;=\;\cong_{k+1}$ then $\cong_{k+1}\;=\;\cong_{k+2}$.
This follows from \eqref{lon} and
$\cong_{k+1}\;\subseteq\;\cong_{k+2}$. Suppose the latter is
not true,
then $\pi_{k+1}(P_i)=\pi_{k+1}(P_j)$ while
$\pi_{k+2}(P_i)\neq\pi_{k+2}(P_j)$. The only possible
cases are that $P_i=\pcc a{P_m}{P_l}$ and $P_j
=\pcc a{P_{m'}}{P_{l'}}$ and 
$\pi_{k+1}(P_m)\neq \pi_{k+1}(P_{m'})$ or 
$\pi_{k+1}(P_l)\neq \pi_{k+1}(P_{l'})$.
So by $\cong_k\;=\;\cong_{k+1}$, at least one of 
$\pi_{k}(P_m)\neq \pi_{k}(P_{m'})$ and 
$\pi_{k}(P_l)\neq \pi_{k}(P_{l'})$ must be true, 
but this refutes
$\pi_{k+1}(P_i)=\pi_{k+1}(P_j)$. So, once the sequence
\eqref{lon} becomes constant, it remains constant.
Since this sequence is decreasing and the maximum 
number of equivalence classes on $\{P_1,\ldots,P_n\}$
is $n$, at most the first
$n$ relations in the sequence can be unequal, hence
$\cong_{n-1}\;=\;\cong_n$, and thus $\pi_{n-1}(P_i)=
\pi_{n-1}(P_j)$ implies $\pi_k(P_i)=\pi_k(P_j)$ for all $k\in\nat$.

It is not difficult to show for threads 
$P$ and $Q$: if $\pi_k(P)=\pi_k(Q)$ for all $k\in\nat$
then $P=Q$. First, each (infinite) thread is
a projective sequence on which $\pi_k$
is defined componentwise. Secondly, for a projective
sequence
$(P_n)_{n\in\Nat}$ it follows that 
$\pi_k(P_k)=\pi_k(\pi_k(P_{k+1})=\pi_k(P_{k+1})=P_k$ for all 
$k\in\nat$. So, for $(Q_n)_{n\in\Nat}$ a projective sequence,
$P_k=\pi_k(P_k)=\pi_k(Q)=Q_k$ for all $k$
implies $(P_n)_{n\in\Nat}=(Q_n)_{n\in\Nat}$.

\section{Thread-Service Composition}
\label{app:C}
Most of this text is taken from~\cite{PZ06}.
A \emph{service}, or a \emph{state machine}, 
is a pair $\langle\Sigma, F\rangle$
consisting of a set $\Sigma$ of so-called \emph{co-actions}
and a reply function $F$.
The reply function is a mapping that gives for each non-empty
finite sequence of co-actions from $\Sigma$ a reply \tr\ or \fa.  

\begin{example}\label{service}
  A \emph{stack} can be defined as a service 
  with co-actions 
  $\method{push{:}i}$,
  $\method{topeq{:}i}$, and
  $\method{pop}$, 
  for $i=1,\ldots,n$ for some $n$, where
  \method{push{:}i} pushes $i$ onto the stack
  and yields \tr,
  the action \method{topeq{:}i} tests whether $i$ is on top 
  of the stack, and
  \method{pop} pops the stack with 
  reply \tr\ if it is non-empty, 
  and it yields \fa\ otherwise.
\end{example}

Services model (part of) the execution
environment of threads. In order to define the interaction 
between a thread and a service,
we let actions be of the form \fma{c}{m}
where \focus{c} is the so-called \emph{channel} or \emph{focus},
and \method{m} is the co-action or \emph{method}.
For example, we write \fma{s}{pop} to denote the action which 
pops a stack via channel \focus{s}.
For service $\ser=\langle\Sigma, F\rangle$ and %finite
thread $P$, 
$P\use{c}\ser$ represents $P$ \emph{using the service} 
\ser\ via channel \focus{c}.
The defining rules for threads in \BTA\ are:
\begin{align*}
  \st\use{c}\ser &= \st,\\
  \di\use{c}\ser &= \di, \\ \displaybreak[0]
  (\pcc{\fma{c'}{m}}PQ)\use{c}\ser &=
 \pcc{\fma{c'}{m}}{(P\use{c}\ser)}{(Q\use{c}\ser)}
 \quad\text{if } \focus{c'}\neq \focus{c},\\
  (\pcc{\fma{c}{m}}PQ)\use{c}\ser &=
 P\use{c}\ser'
 \quad\text{if }\method{m}\in\Sigma\text{ and }F(\method{m})=\tr,
 \\\displaybreak[0]
  (\pcc{\fma{c}{m}}PQ)\use{c}\ser &=
 Q\use{c}\ser'
 \quad\text{if }\method{m}\in\Sigma\text{ and }F(\method{m})=\fa,\\
  (\pcc{\fma{c}{m}}PQ)\use{c}\ser &= \di
 \quad\text{if }\method{m}\not\in\Sigma,
\end{align*}
where $\ser'= \langle\Sigma, F'\rangle$ with 
$F'(\sigma) = F(\method{m}\sigma)$ 
for all co-action sequences $\sigma\in\Sigma^+$.

The operator $\use c$ is called
the \emph{use operator} and stems from~\cite{BP02}.
An expression $P\use{c}\ser$ is
sometimes referred to as a \emph{thread-service composition}.
The use operator is expanded to infinite threads 
in \BTAinf\ by defining
\[
  (P_n)_{n\in\nat}\use c \ser = 
  \bigsqcup_{n\in \nat}P_n\use c \ser.
\]
(Cf.\ \cite{BB05}.) It follows that the rules
for finite threads are valid for infinite
threads as well.
Observe that the requests to the service
do not occur as actions in
the behavior of a thread-service composition. 
So the composition not
only reduces the above-mentioned
non-determinism of the thread,
but also hides the associated actions.

In the next example we show
that the use of services may turn regular threads 
into non-regular ones.
\begin{example}\label{example:irregpd}
We define a thread using a stack as defined in Example~\ref{service}.
We only push the value $1$ (so the stack behaves as a counter),
and write $S(n)$ for a stack holding $n$ times the value $1$.
By the defining equations for the use operator it follows that
for any thread $P$,
\begin{align*}
 (\fma{s}{push{:}1}\circ P)\use s S(n) &=
 P\use s S(n{+}1),\\
 (\pcc{\fma{s}{pop}}P\st)\use s S(0) &= \st,\\
 (\pcc{\fma{s}{pop}}P\st)\use s S(n{+}1) &= P\use s S(n).
\end{align*}
Now consider the regular thread $Q$
defined by
\[
  Q = \pcc a{\fma{s}{push{:}1}\circ Q} R,\quad
  R = \pcc{\fma{s}{pop}}{b\circ R}\st,
\]
where actions $a$ and $b$ do not use focus \focus{s}.
Then, for all $n\in\nat$,
\begin{align*}
  Q\use s S(n) &=
 (\pcc a{\fma{s}{push{:}1}\circ Q}R)\use s S(n)\\
  {} &= \pcc a{(Q\use s S(n{+}1))}{(R\use s S(n))}.
\end{align*}
It is not hard to see that $Q\use s S(0)$
is an infinite thread with the property that for all $n$,
a trace of $n+1$ $a$-actions produced by $n$ positive
and one negative reply on $a$ is followed by $b^n\circ\st$.
This yields an \emph{non-regular} thread:
if $Q\use s S(0)$
were regular, it would be a fixed point of some finite linear
recursive specification, say with $k$ equations.
But specifying a trace $b^k\circ\st$ already requires $k+1$
linear equations 
$x_{1}=b\circ x_{2},\dots,x_{k}=b\circ x_{k+1},x_{k+1}=\st$, 
which contradicts the assumption.
So $Q\use s S(0)$ is not regular.
\end{example}

Finally, we note that the use of finite state services,
such as Boolean registers, can \emph{not} turn regular
threads into non-regular ones (see~\cite{BP02}).
More information on thread-service composition can be
found in e.g.~\cite{PZ06}.

\end{document}